# Two phase helium cooling characteristics in Cable-in Conduit Conductors

**G.K. Singh, S. Pradhan and V.L. Tanna**

**Institute for Plasma Research, HBNI, Bhat, Gandhinagar-382 428, Gujarat, India.**

**E-mail: gaurav.singh@ipr.res.in , pradhan@ipr.res.in, vipul@ipr.res.in**

## Abstract

Cable-in-Conduit Conductors (CICCs) are used in the fabrication of superconducting fusion grade magnets. It acts as a narrow cryostat to provide cryo-stability with direct contact of coolant fluid to conductor. The superconducting magnets are cooled using forced flow (FF), supercritical helium or two phase (TP) cooling through void space in the CICC. Thermo-hydraulics using supercritical helium single phase flow is well-known and established. Research topic of behavior of forced flow, two phase (TP) helium cooling in CICC involves perceived risks of the CICC running into flow chocking and possible thermo-acoustic oscillations leading to flow instabilities. This research work involves study of forced flow two phase helium cooling in CICC wound superconducting magnets. The TP flow provides cryo-stability by the latent heat of helium not by enthalpy as in case of CICC being cooled with supercritical helium. Study reveals some attractive regimes in the case of TP cooling, at a given mass flow rate of single phase helium at the inlet and a heat flux acting on the CICC. Analysis carried out predicts significant gains with TP cooling on a prototype CICC, which is circular in cross section and appropriate for fusion devices for high magnetic field applications. These general formalisms may be extended to specific magnets wound with CICC. This paper describes analysis of TP cooling of a CICC.

## Introduction

Cable-in-Conduit Conductors (CICCs) are attractive options in fabrication of superconducting fusion grade magnets. CICC architecture provides better mechanical stability, greater wetted perimeters to the twisted strands and overall protection to the superconducting wires. A CICC configuration is often considered as a narrow cryostat providing adequate cryo-stability to the cable and in it the conductor is in direct contact with the flowing fluid (coolant). As helium flows through a conductor, there is a pressure drop across the flow path.

Liquid helium two phase flow is commonly distributed near saturation conditions. The flow behavior and thermo-hydraulic problems become complex in TP as there are several variables that affect it such as mass flow rate, phase distribution, pressure and temperature relative to saturation, heat transfer, rate of phase change etc. Description of two models explaining two phase flow can be found in literature [1]. However, much of this work has been carried out on conventional fluids (water/steam, water/air). There are few analytical works done for cryogenic two phase flow [2-5]. Several risks associated with two phase flow such as chocking and flow instabilities [6-8] on account of which the TP cooling is not widely preferred over the single phase cooling.

Advantage of two phase flow is the heat removal capability. Heat transfer in the TP cooling is higher as compared to the single phase flow, as it utilises its latent heat. The possibility of cooling CICC wound superconducting magnets by two phase helium flow in this research work is studied on a prototype CICC design. Thermo-hydraulic behaviour in TP cooled CICC wound magnets critically seeks information on the vapor quality, heat flux on superconductors, and temperature distribution along the superconducting magnet. This work aims at quantitative analysis of some of this critical information in TP cooling. The work involves study of the pressure drop, effective temperature and outlet vapor quality of two-phase flow over long steady state operations in a typical CICC wound magnet.

As helium flows through cooling channel, due to pressure drop and static heat flux, vaporization takes place by means of utilizing latent heat resulting in possible two phase flow. This phenomenon in a two phase flow of helium in a cooling channel is shown in figure 1. Mass flow rate measurement is not accurately possible in case of TP cooling. Due to lack of both theoretical models and reliable experimental data on these issues, the present work is aimed at analyzing qualitatively the thermo-hydraulic characteristics of such a channel, and approximately evaluates the fluid resistance and other parameters of the flow.

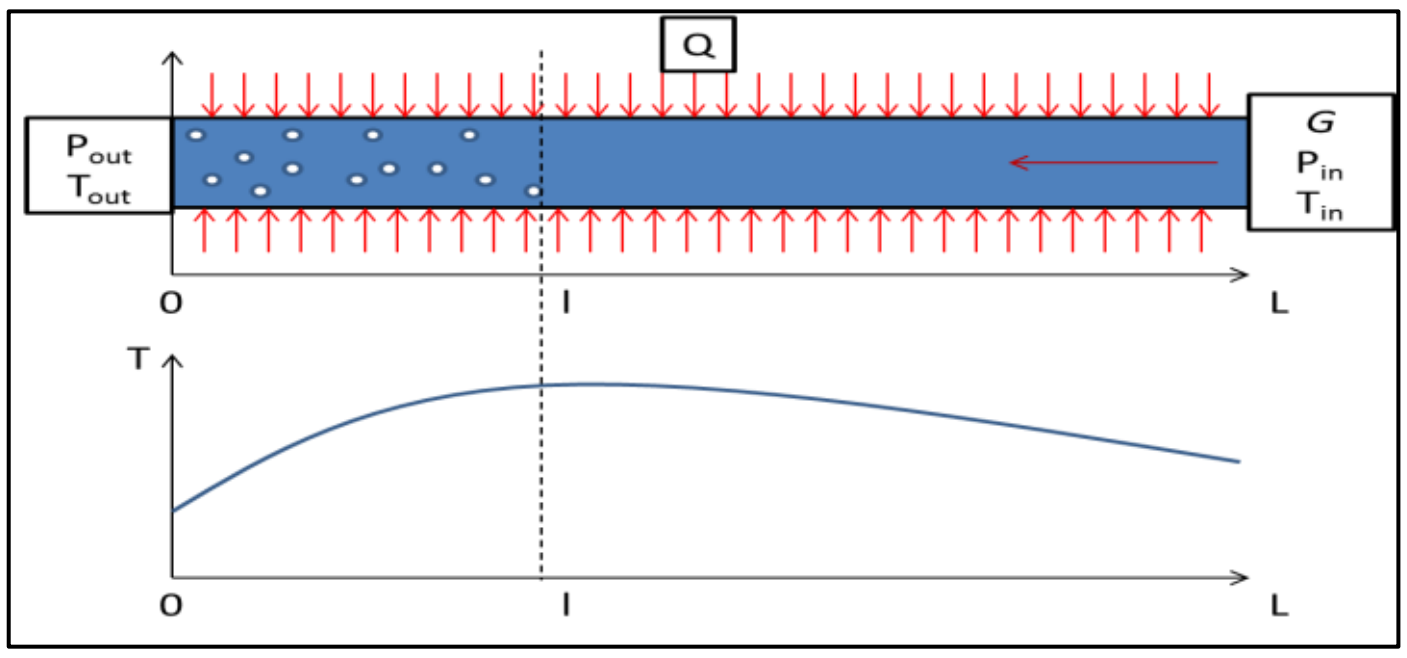

Figure 1:1-D representation of CICC and temperature distribution along it in the case of the two phase flow

As shown in figure1, stable single phase sub-cooled LHe is introduced at inlet of a horizontally heated CICC (Specification of CICC given in Table 2). The magnets are thermally well insulated and the wall temperatures are mostly constant along the length of flow. As LHe moves in the channel, it gets heated up, and part of liquid content is converted into vapor. At certain cross-section of CICC, LHe temperature reaches saturation temperature and starts boiling. The flow in the channel beyond this point is in two phase. Vapor quality rises as a result of heating along the length of CICC. It is necessary for us to find the fluid hydraulic resistance, i.e., the dependence of the pressure drop on the flow rate of the liquid, and the maximum temperature of the liquid along the channel.

In the pressure range of 1.0-2.3 *bar* (a) corresponding to its saturation temperatures, the fluid exhibits two-phase flow behaviour when heat is added. In two phase mode, temperature of helium does not change but the quality changes through latent heat of vaporization. Pressure drop increases as mass flow rate increases and, it depends on frictional force acting on its flow through channel length. Under this study, these considerations are extended to a CICC cooled with TP helium following the analytical studies as explained in [9].

**Formulae for CICC Two phase Pressure drop and effective temperature [9]**

$$\Delta P = \left(\psi \frac{m}{\beta} + \frac{sQ}{\dot{m}b}\right)\left[1 - \frac{1}{m}\ln\left(1 + \frac{x_0 n}{1+\frac{x_0 n}{m}}\right)\right] \quad (1)$$

$$\psi = 1 + \varphi \frac{2d}{f(L-l)} \quad (2)$$

$$m = \varepsilon \beta v' G^2 L \quad (3)$$

$$s = 1 + (h'_0 - h_{in})\dot{m}/Q \quad (4)$$

$$\varepsilon = \frac{f}{2d} \quad (5)$$

$$\beta = \frac{bn}{L_v} \quad (6)$$

$$x_0 = qLs/\dot{m}L_v \quad (7)$$

$$n = \frac{v'' - v'}{v'} \quad (8)$$

$$\frac{l}{L} = \frac{1}{m}\ln\left(1 + \frac{x_0 n}{1+\frac{x_0 n}{m}}\right) \quad (9)$$

$$Q = qL \quad (10)$$

$$\dot{m} = GA \quad (11)$$

$$T_{h\,max} = T_{in} + \left(1 - \frac{l}{L}\right)\left(\frac{x_0 L_v}{c_p}\right) \quad (12)$$

Meaning of the symbols used above has been listed under nomenclature on page number 7. In this analytical solution, following assumptions have been considered: Liquid flow is homogenous in Two-Phase section, Thermo-physical properties of liquid and vapor are constant, Liquid is incompressible, Flow velocity is small and, hence there is no additional acceleration induced pressure drop, Heat transfer coefficient from wall to liquid is infinitely large i.e. liquid and wall has same temperature, thermal flux is independent of wall temperature and the enthalpy of liquid on saturation line is independent of pressure.

With above assumptions, the present analytical study of on CICC being cooled with TP helium has been carried out. The effective temperature , vapour content at the outlet, and pressure drop for helium flow as a function of various mass flow rates for various heat flux have been estimated for prototype CICC. The prototype CICC specifications have been elaborated in Table 2. Table-1 below gives the gross input parameters for analysis of TP flow in such a CICC.

Table 1: CICC hydraulics parameters & Input parameters

| **Parameters** | **Unit** | **Prototype CICC** |
|---|---|---|
| Outlet Pressure | bar (a) | ~1.4 |
| Outlet Temperature | K | 4.6 |
| Heat Flux | W/Path | 4 – 6 |
| Mass flow rate | g/s | 0 – 1 |

**Case of the prototype superconducting CICC**

The prototype CICC is circular in cross section as shown in fig-2. The strands could be high current carrying high field NbTi or $Nb_3Sn$ or $Nb_3Al$ or $MgB_2$. The strands are cabled in a twisting scheme of 3×3×3×5 configuration. The final twisted cable is compacted and wrapped inside thin SS 316 LN foil before being pulled through inside a non-magnetic SS 316 LN conduit of inner diameter 13.6 mm and wall thickness of 1.7 mm. The CICC has a void fraction (ν) of 40 %±2 %. (Specification given in Table 2)

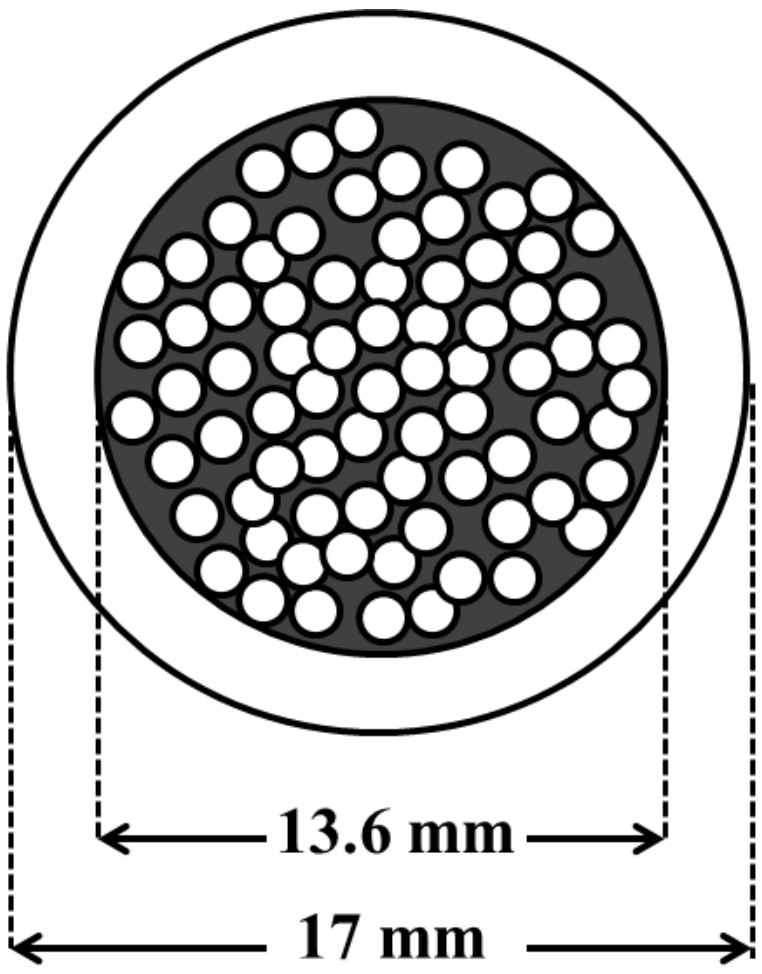

**Figure 2:** Schematic cross sectional view of the prototype CICC

For ~~the~~ analysis, ~~th~~e typical operating pressure of TP helium at 1.6 bar (a) is adopted and operating temperature is the corresponding saturation temperature of helium for 50-m long CICC test section. The inlet to CICC is pure liquid helium being heat exchanged in a sub-cooler Dewar as has been the practice in large devices. Using the assumption of homogenous as well as considering that fluid is in pure liquid phase at the inlet, the vapor quality evolution at the outlet ($x_o$) is to be determined. In TP cooled cases, the density and viscosity of the two phase mixture cannot be predicted without knowing the vapour void (α) and quality factor ($x$).

Table 2: Specifications of prototype CICC

| | Unit | Value |
|---|---|---|
| Test path length | m | 50 |
| Outer Diameter | m | 1.7E-02 |
| Inner Diameter | m | 1.36E-02 |
| Diameter of Strand, $D_{st}$ | m | 9.00E-04 |
| Total Area , $A_t$ | $m^2$ | 1.45E-04 |
| Flow area of LHe, $A_{he}$ | $m^2$ | 5.8E-05 |
| Wetted Perimeter, $P_{cool}$ | m | 3.21E-01 |
| Hydraulic Diameter, $D_h$ | m | 7.23E-04 |

## Analytical Results and Discussion

Using the inputs elaborated in table-2, and adopting the formalism explained in equation 1-12, the following predictions have been made: (i) Effective temperature vs. the mass flow rate for a fixed hydraulic length and constant heat flux. (Figure 3) (ii) Changes in the vapor fraction (`figure of merit' of two-phase) with mass flow rate for fixed hydraulic length and constant heat flux. (Figure 4) (iii) Pressure drop and outlet temperature across a fixed hydraulic path with constant lead flux as a function of mass flow. In the present work, analysis has been carried out for four different realistic heat flux of 0.08 W/m, 0.10 W/m, 0.11 W/m and 0.12 W/m for total path length of 50 m having nominal flow of 0.33±0.02 g/s (Figure 5)

Table 3: Hydraulic analysis of typical CICC

| Heat flux (W/m) | Path length (m) | Total heat flux/Path (W) | Mass flow/Path (g/s) | Outlet vapor quality | Inlet Pressure (bar(a)) |
|---|---|---|---|---|---|
| 0.08 | 50 | 4.0 | 0.31 | 0.68 | 1.53 |
| | | | 0.33 | 0.64 | 1.55 |
| | | | 0.35 | 0.60 | 1.57 |
| 0.10 | 50 | 5.0 | 0.31 | 0.85 | 1.54 |
| | | | 0.33 | 0.79 | 1.56 |
| | | | 0.35 | 0.75 | 1.58 |
| 0.11 | 50 | 5.5 | 0.31 | 0.93 | 1.55 |
| | | | 0.33 | 0.88 | 1.56 |
| | | | 0.35 | 0.83 | 1.58 |
| 0.12 | 50 | 6.0 | 0.31 | 0.98 | 1.55 |
| | | | 0.33 | 0.95 | 1.57 |
| | | | 0.35 | 0.90 | 1.59 |

The sub-cooled liquid boils as it moves along the channel since the pressure reduces along the channel length. Subsequently, the fluid temperature reaches to its saturation temperature in a certain cross-section of the channel. Increase in the flow rate of helium results in reduction in vapor content in outlet and reduction in the length of two phase section. At given mass flow rate, the effective temperature is maximum. Effective temperature is less pronounced for lower heat flux. Pressure drop depends on vapor quality and mass flow for various heat fluxes. At a given mass flow and uniform heat flux; pressure drop, effective temperature and vapour quality have been analysed.

These analytical results and benefits of the two-phase cooling may be verified, if such experiments get conducted. These results show that CICC wound superconducting magnets can be operated in two phase mode in a defined operating regime in safe and reliable way. For a fixed path length, in order to lower the effective temperature one can operate in the region of relatively low mass flow rates and high vapor quality at outlet. This will increase the two phase section of the path and it enhances the heat transfer in higher vapor content region.

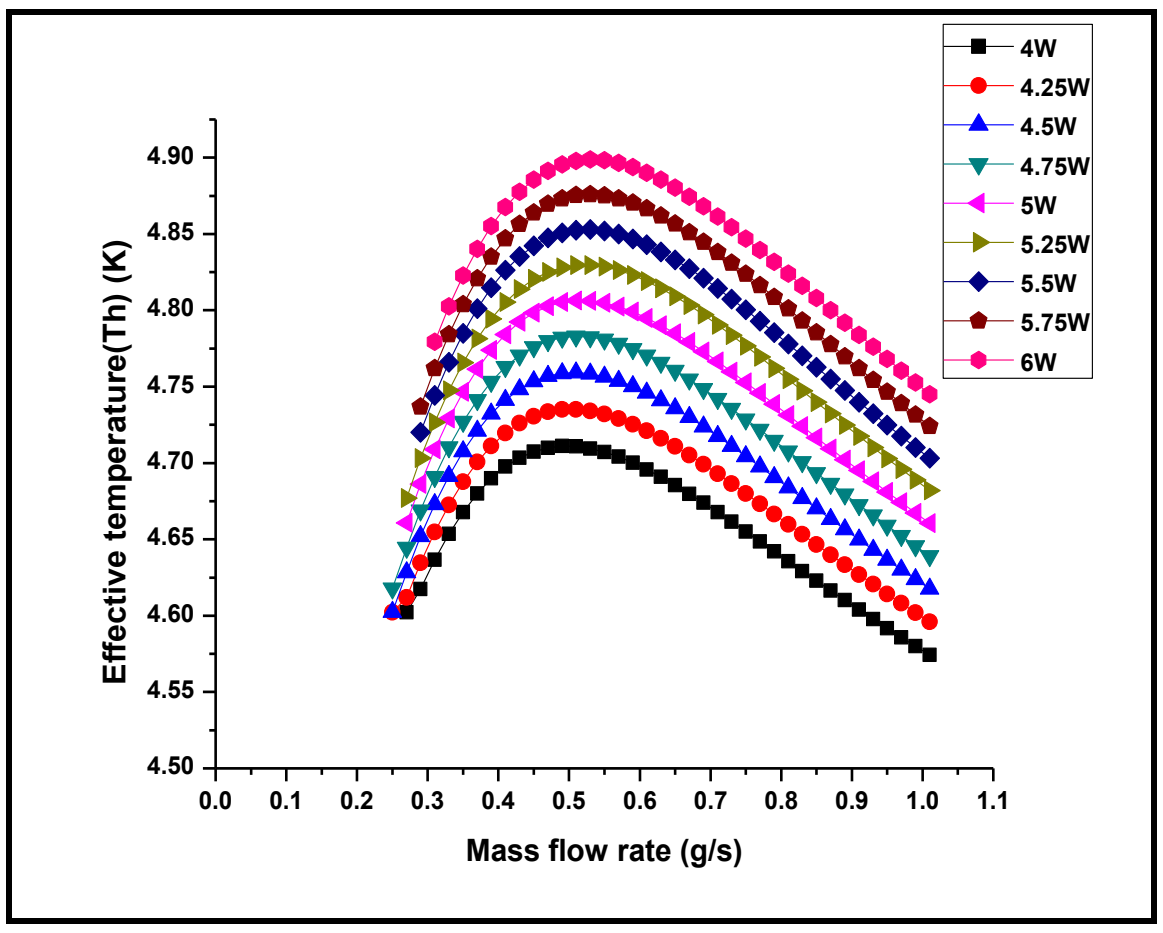

Figure 3: Effective Temperature variance at different heat flux for prototype CICC

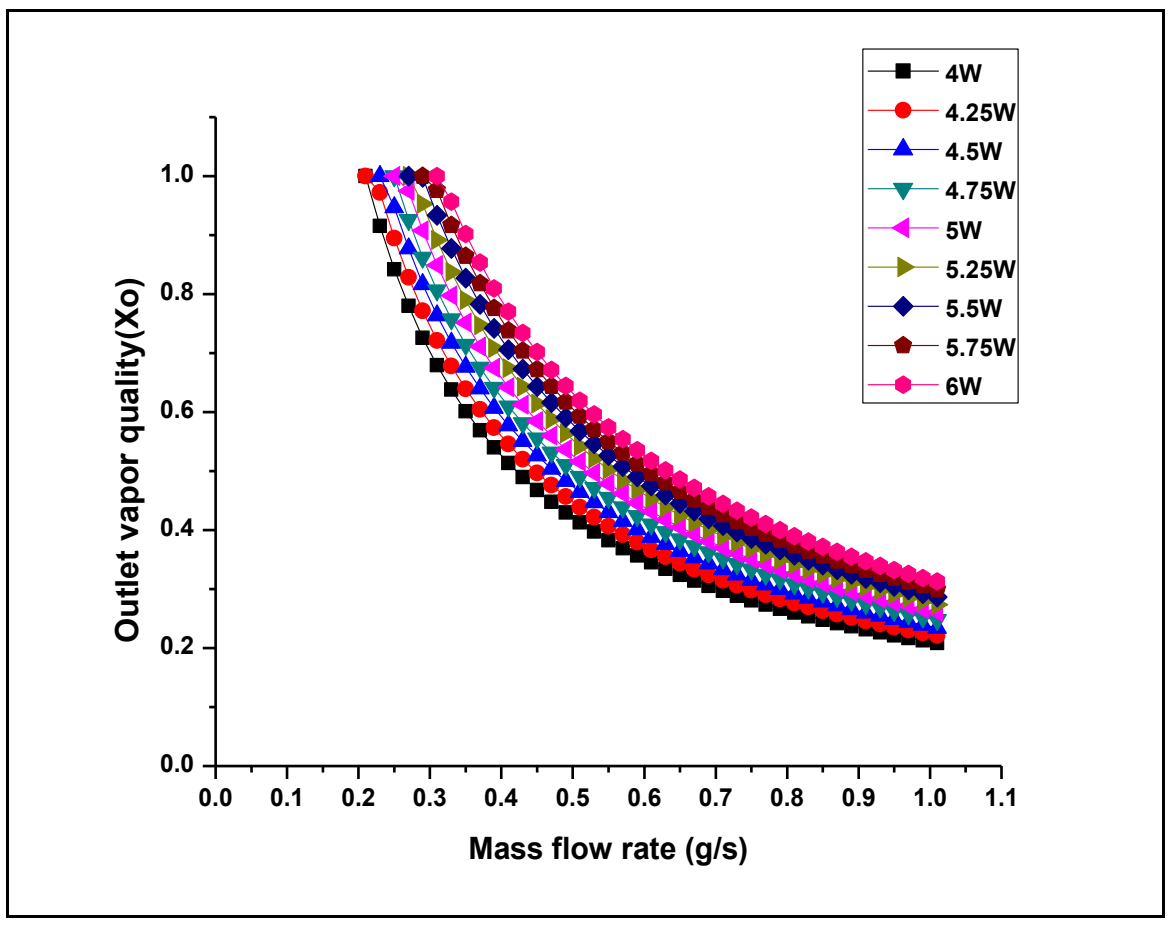

Figure 4: Vapor Content variance at different heat flux for prototype CICC

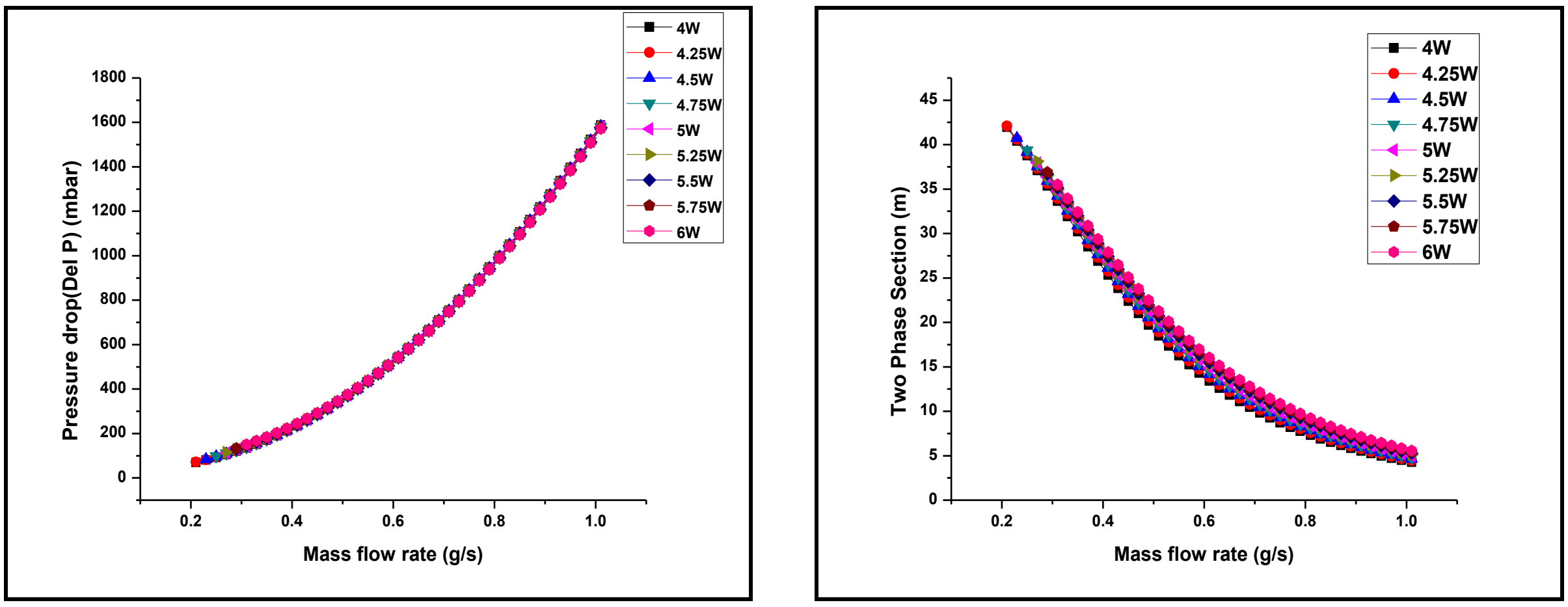

Figure 5: Pressure Drop variance at different heat flux for prototype CICC

## Conclusions

The pressure drop and quality factor analysis has been carried for a prototype CICC wound high field superconducting magnet for a number of realistic heat flux and inlet mass flow rates. The effective temperatures have also been predicted. The analysis may be useful for future two phase flow related experiments in a complex geometry like CICC. The results also state that as the mass flow rate increases, corresponding pressure drop increases and hence the vapor quality decreases for a given heat flux. Analysis have shown results that CICC wound fusion magnets may be operated in the two phase flow of helium under certain operational envelopes as discussed. The two-phase cooling of CICC do get into risks with possibilities of flow choking and thermo-acrostic instabilities, but ensuring single phase sub-cooling at the inlet these possibilities in practice could be reduced significantly for certain heat flux and operating parameters. The obvious advantage of the two-phase cooling as against the single phase cooling is the reduced requirements of the mass flow rates. Thus, in a cryogenic system, the stringent requirements of a cold circulator and its associated heat flux budget may be eliminated or at least reduced. Further, the resulting cooling scheme becomes simpler. Even though, the TP cooling and its thermo-hydraulics is complex in nature to realize, by using the prescriptions discussed, reasonable predications of these quantities are feasible (effective temperature, pressure drop and vapour quality). This information may be helpful in practical operations of TP cooled magnets.

## Nomenclature

$\Delta P = Pressure\ drop$
$Q = Heat\ load$
$\dot{m} = Mass\ flow\ rate$
$G$ = Mass Flux (kg/s.$m^2$)
$A$ = Helium cross-sectional area ($m^2$)
$Void$ = 0.4
$d$ = Hydraulic diameter (m)
$x_o = Vapor\ quality$
$\varphi = coefficient\ of\ fluid\ resistance\ in\ the\ single\ phase\ section$
$m = Dimensionless\ Index$
$f = Friction\ coefficient$
$R_e = Reynold's\ Number$
$b = 0.59\ j.\frac{m^2}{kg.N}$
$L_v =\ Latent\ heat\ of\ vaporization\ (kj/kg)$
$v' = Specific\ volume\ of\ Liquid$
$v'' =\ Specific\ volume\ of\ vapor$
$x_0 = mass\ content\ of\ vapor\ at\ output$
$l = Two - Phase\ flow\ length (m)$
$L =\ Length\ of\ Channel\ (m)$
$C_p = Isobaric\ Heat\ Capacity\ (kj/kg.K)$
$h_{out} = Enthalpy\ at\ output$
$h_{in} = Enthalpy\ at\ input$
$h'_0 = liquid\ Enthalpy\ on\ the\ saturation\ line$
$T_h = Effective\ Temperature\ (K)$